\documentstyle[preprint,aps,prl]{revtex}
\begin{document}
\draft
\preprint{SNUTP  95--064}
\title{Inclusive $S-$wave charmonium \\
productions  in $B$ decays   }
\author{ Pyungwon Ko$^{1,}$\footnote{ pko@phyb.snu.ac.kr}}
\address{$^{1}$Department of Physics, Hong-Ik University,
Seoul 121-791, Korea }
\author{Jungil Lee$^{2,}$\footnote{jungil@phyy.snu.ac.kr}
    and H.S. Song$^{2,}$\footnote{hssong@phyy.snu.ac.kr} }
\address{
$^{2}$Department of physics and Center for Theoretical Physics
\\ Seoul National University,
Seoul 151-742, Korea  }
\date{\today}
\maketitle
 \tighten
\begin{abstract}
The inclusive $S-$wave charmonium production rates  in $B$ decays
are considered
using the Bodwin-Braaten-Lepage (BBL) approach, including the
relativistic corrections and  the color-octet mechanism suggested
as a possible solution to the $\psi^{'}$ puzzle at the Tevatron.
We first consider relativistic
and radiative corrections to $J/\psi \rightarrow e^{+} e^{-}$ and
$J/\psi \rightarrow$ Light Hadrons (LH), in order
to determine two nonperturbative parameters, $\langle J/\psi | O_{1}
(^{3}S_{1}) | J/\psi \rangle$, $\langle J/\psi | P_{1} (^{3}S_{1})
| J/\psi \rangle$,
in the factorization formulae for these decays.
Using these two matrix elements and including the color-octet
$c\bar{c}(^{3}S_{1})$ state contribution,
we get a moderate increase in the decay
rates for $B$ decays into $J/\psi ~({\rm or}~\psi^{'}) ~+ X$.
Our results, $B(B \rightarrow J/\psi~({\rm or}~\psi^{'}) + X)
= 0.58~(0.23) \%$
for $M_{b} = 5.3$ GeV, get closer to the recent CLEO data.
As a byproduct, we prefer a larger decay rate for $\eta_{c} \rightarrow$
LH  compared to the present data.
\end{abstract}


\narrowtext

 \section{Introduction}
 \label{sec:intro}

It has been commonly believed that inclusive production rates of a
heavy quarkonium state in various
high energy processes can be  adequately described
by perturbative QCD (PQCD) within the color-singlet model
\cite{pqcd}.
However, recent observations of inclusive $J/\psi$ and $\psi^{'}$
productions at high $p_T$ at the Tevatron suggest that a new mechanism
is called for beyond the color-singlet model \cite{cdf}.  For the
$J/\psi$ production, the lowest order subprocess comes from
parton fusions, which are smaller than the data by more than an order of
magnitude.  If one includes the gluon fragmentations into $J/\psi$ and into
$\chi_{cJ}(1P)$ states followed by $\chi_{cJ}(1P) \rightarrow J/\psi +
\gamma$ (which is the next-to-leading order in $\alpha_s$), one gets higher
theoretical estimates that still underestimates the experimental yield by
factor of $3 \sim 5$.  For the $\psi^{'}$  production, the situation
is even worse.  Even with the gluon fragmentation included, the theoretical
production rate falls below the data by factor of $\sim 30$ or so.

In  order to resolve this puzzle, basically two scenarios have been
suggested up to now : (i) existence of new charmonium states above the
$D \bar{D}$ threshold, which can decay into $J/\psi$ and $\psi^{'}$ with
appreciable branching ratios \cite{wise}--\cite{roy}, and (ii) importance
of  the gluon fragmentation
into a pointlike color-octet $S-$wave $c \bar{c}$ ($^{3}S_{1}$) state,
and its subsequent evolution into $\psi^{'}$ \cite{braaten}.
Both scenarios are quite intriguing in a
sense that they call for new elements of physics within the standard model,
{\it new spectroscopy } or {\it new production mechanism}  for charmonium
states.  It would be useful to explore and test these suggestions in places
other than $p \bar{p}$ colliders such as the Tevatron.
In Ref.~\cite{ko}, one of
us has explored the
consequences of the first scenario, finding that these hypothetical
$\chi_{cJ}(2P)$ states should be observed at the level of $(0.3-0.5) \%$
(in branching ratio) in the decay channels of $B \rightarrow
\chi_{c,J=1,2}(2P) + X$
(followed by $\chi_{c,J=1,2} \rightarrow \psi^{'} + \gamma$ with
$\approx 10 \%$ in branching ratio, if the first scenario is to work).

In this paper, we explore the consequences of the second scenario in the
inclusive decays of $B$ mesons into a $S-$wave charmonium.
It is well known that
the lowest order results in the heavy quark velocity and the strong coupling
constant for $\Gamma ( B \rightarrow J/\psi + X)_{\rm direct}$ is smaller
than the CLEO data by factor of $\sim 3$ \cite{bodwinB}.
The situation does not get better even if the next-to-leading order
corrections in $ \alpha_s$ are included in the nonleptonic effective weak
hamiltonian for $B$ decays \cite{bergstrom}.
In Ref.~\cite{bodwinB}, only the
color-singlet contribution has been included, since the color-octet
contributions are higher order in $v^2$, hence suppressed relative to
the singlet contributions. However, the Wilson coefficient for the
color-singlet contribution is much suppressed
compared to that for the color-octet
contribution.  Therefore, the color-octet contribution, being suppressed by
$v^4$, may be numerically important, because of the larger Wilson
coefficient. This is similar to the case of the gluon fragmentation into
$\chi_{cJ}(1P)$  states \cite{gtochi},
for which the color-octet contribution is lower order in $\alpha_s$
compared to the color-singlet contribution, whereas both of them are of the
same order in $v^2$. Also, in the case of $\psi^{'}$ production through
the gluon fragmentation into a color-octet $c\bar{c}$ state, the octet
contribution is higher by $v^4$ compared to the color-singlet contribution,
but this is compensated by the large short distance factor $1/\alpha_s^2$
compared to the color-singlet contribution \cite{braaten}.

In Sec.~\ref{sec:two}, the $S-$wave charmonium production rates in $B$
decays are calculated in the framework of Nonrelativistic QCD (NRQCD)
\cite{bodwin} including the relativistic
corrections and a color-octet $[c\bar{c}(^{3}S_{1})]$ contribution which is
of $O(v^{4})$ compared to the nonrelativistic limit.  The results contain
three nonperturbative parameters, $\langle 0 | O_{1}^{H} (^{3}S_{1})
| 0 \rangle, \langle 0 | P_{1}^{H} (^{3}S_{1}) | 0 \rangle$ and $\langle 0
| O_{8}^{H} (^{3}S_{1}) | 0 \rangle$.  Among these three parameters
appearing in the heavy quarkonium productions,
the first two color-singlet matrix elements can be related another
parameters,  $\langle H | O_{1} (^{3}S_{1}) | H \rangle $ and
$\langle H | P_{1} (^{3}S_{1}) | H \rangle $, which
enters in the heavy quarkonium decays in the vacuum saturation
approximation \cite{bodwin}, via
\begin{equation}
\langle 0 | O_{1}^{H} | 0 \rangle \approx (2J+1)~\langle H | O_{1}
| H \rangle~\left( 1 + O(v^{4}) \right).
\end{equation}
In Sec.~\ref{subsec:31}, the two parameters $\langle H | O_{1} (^{3}S_{1})
| H \rangle$ and $\langle H | P_{1} (^{3}S_{1}) | H \rangle$ (with $H =
J/\psi, \psi^{'}$)   are determined
by analyzing the decays of $J/\psi$ and $\psi^{'}$ into light hadrons
(LH) and  $e^{+} e^{-}$.
Implications of this analysis on the decays of $\eta_c$ into light hadrons
and $\gamma \gamma$ are discussed.  Our result prefers larger decay rate for
$\eta_{c} \rightarrow $ LH.  In Sec.~\ref{subsec:32}, we present numerical
estimates for $S-$wave charmonium production rates in $B$ decays, using the
factorization formulae obtained in Sec.~\ref{sec:two} and $\langle 0
| O_{1}^{H} (^{3}S_{1})
| 0 \rangle, \langle 0 | P_{1}^{H} (^{3}S_{1}) | 0 \rangle$ obtained
in Sec.~\ref{subsec:31}. We also discuss the polarization of the $J/\psi$
in $B$ decays.  In Sec.~\ref{sec:con}, the results are summarized, and
possible improvements of the present work are speculated.

 \section{Higher order corrections}
 \label{sec:two}

The effective Hamiltonian for $b \rightarrow c \bar{c} q$ (
with $q = d,s$) is written as \cite{bodwinB}
\begin{eqnarray}
H_{eff} & = & {G_{F} \over \sqrt{2}}~V_{cb} V_{cq}^{*}~
\left[ {{2 C_{+} - C_{-}}
\over 3}~\bar{c} \gamma_{\mu} (1 -\gamma_{5})c~\bar{q} \gamma^{\mu} (1 -
\gamma_{5}) b ~ \right.
\nonumber    \\
 & + & \left. (C_{+} + C_{-})~\bar{c} \gamma_{\mu} (1 - \gamma_{5})
T^{a} c ~\bar{q}\gamma^{\mu} (1 - \gamma_{5}) T^{a} b \right],
\label{eq:heff}
\end{eqnarray}
where $C_{\pm}$'s are the Wilson coefficients at the scale $\mu \approx
M_b$. We have neglected penguin operators, since their Wilson coefficients
are small  and thus they are irrelevant to our case.
To leading order in $\alpha_{s}(M_{b})$ and to all orders in $\alpha_{s}
(M_{b})~{\rm ln}(M_{W}/M_{b})$,  the above Wilson coefficients are
\begin{equation}
C_{+} (M_{b}) \approx 0.87,~~~~~ C_{-} (M_{b}) \approx 1.34.
\end{equation}

According to the factorization theorem for the $S-$wave charmonium
productions in  $B$ decays,  one has
\cite{bodwinB}
\begin{eqnarray}
\Gamma ( b \rightarrow J/\psi + X )  =   { \langle 0 | O_{1}^{J/\psi}
(^{3}S_{1}) | 0 \rangle \over 3 M_{c}^2}~ \hat{\Gamma}_{1}
( b \rightarrow (c \bar{c})_{1} (^{3}S_{1}) + X ),
\\
\Gamma ( b \rightarrow \eta_{c} + X )  = { \langle 0 | O_{1}^{\eta_c}
 (^{1}S_{0})
 | 0 \rangle \over M_{c}^2}~\hat{\Gamma}_{1}
( b \rightarrow (c \bar{c})_{1} (^{1}S_{0}) + X ),
\end{eqnarray}
in the nonrelativistic limit,
where $\hat{\Gamma}_{1}$ are rates for hard subprocesses of $b$ quark
decaying into a $c \bar{c}$ pair with suitable angular momentum and
vanishing  relative momentum in the color-singlet :
\begin{eqnarray}
\hat{\Gamma}_{1} (b \rightarrow (c\bar{c})_{1}
(^{3}S_{1}) + s,d) & = & (2C_{+} - C_{-}
)^{2} \left( 1 + {8 M_{c}^{2} \over M_{b}^2} \right)~\hat{\Gamma}_{0},
\\
\hat{\Gamma}_{1} (b \rightarrow (c\bar{c})_{1}
(^{1}S_{0}) + s,d) & = &
(2C_{+} - C_{-} )^{2} ~\hat{\Gamma}_{0},
\end{eqnarray}
with
\begin{equation}
\hat{\Gamma}_{0} \equiv |V_{cb}|^{2} \left( {G_{F}^{2} \over 144 \pi}
\right) M_{b}^{3} M_{c} \left( 1 - {4 M_{c}^{2} \over M_{b}^2} \right)^{2}.
\end{equation}
The operator $O_{1}^{H} (^{2S+1}S_{J})$ is defined in terms of heavy
quark field operators in NRQCD \footnote{We follow the notations in
Ref.~\cite{bodwin}, and will not give explicit forms for these dimension-six
operators in this paper.}.
Its  matrix element $\langle 0 | O_{1}^{H} (^{2S+1}S_{J}) | 0 \rangle$
contains the nonperturbative
effects in the heavy quarkonium production processes,
and is proportional to the probability that a $c \bar{c}$ in a color-singlet
$S-$wave state fragments into a  color-singlet $S-$wave $c \bar{c}$ bound
state such as  a  physical  $J/\psi$, $\eta_c$ or $\psi^{'}$.
It is also  related to the matrix element
$\langle H | O_{1} (^{2S+1}S_{J}) | H \rangle$ as in Eq.~(1), and also
with the nonrelativistic quarkonium wavefunction as follows :
\begin{equation}
\langle 0 | O_{1}^{J/\psi} (^{3}S_{1}) | 0 \rangle \approx
3~\langle J/\psi | O_{1} (^{3}S_{1}) | J/\psi \rangle
\approx \left(
{9 \over 2 \pi} \right)~  \left| R_{\psi} (0)
\right|^2,
\end{equation}
in the nonrelativistic limit. Similar expressions hold for the case of
nonperturbative matrix elements appearing in the $\eta_c$ productions
and its decays :
\begin{equation}
\langle 0 | O_{1}^{\eta_c} (^{1}S_{0})  | 0 \rangle \approx
\langle \eta_{c} | O_{1} (^{1}S_{0})  | \eta_{c} \rangle
\approx \left( {3 \over 2 \pi} \right)~  \left| R_{\eta_c} (0)
\right|^2.
\end{equation}
Note that dependence on the radial quantum numbers
$n$ enters through the nonperturbative parameters, $\langle 0 | O_{1}^{H}
(^{3}S_{1}) | 0 \rangle$.

Using the leptonic decay width of $J/\psi$ and $\psi^{'}$, one can determine
\begin{eqnarray}
\langle J/\psi | O_{1} (^{3}S_{1}) | J/\psi \rangle & \approx &
2.4 \times 10^{-1}~~{\rm GeV}^{3},
\\
\langle  \psi^{'} | O_{1} (^{3}S_{1}) | \psi^{'} \rangle
& \approx & 9.7 \times 10^{-2}~~{\rm GeV}^{3},
\end{eqnarray}
in the nonrelativistic limit with $\alpha_{s} (M_{c}) = 0.27$.
\footnote{The radiative
corrections  in $\alpha_s$ has not been included here for consistency.
To be consistent with the velocity counting rules in the
NRQCD in the Coulomb gauge for the heavy quarkonia \cite{bodwin},
one has to include the relativistic corrections as well, since $v \sim
\alpha_{s} (Mv)$  in heavy quarkonium system. If one includes the
$O(\alpha_{s})$ radiative corrections to $J/\psi \rightarrow l^{+} l^{-}$
without relativistic corrections, one gets a larger $\langle 0 |
O_{1}^{J/\psi} (^{3}S_{1}) | 0 \rangle$ compared to the lowest order
result, Eq.~(11)
: $ \langle 0 | O_{1}^{J/\psi} (^{3}S_{1}) | 0 \rangle \approx
4.14 \times 10^{-1}~~{\rm GeV}^3$.
Relativistic corrections gives a further
enhancement. See Eq.~(48) below.}
Also, to the lowest order in $v^2$, one has $\langle \eta_{c} | O_{1}
 (^{1}S_{0}) | \eta_{c} \rangle = \langle J/\psi | O_{1}
 (^{3}S_{1}) | J/\psi \rangle$
because of heavy quark spin symmetry \cite{bodwin}.  From
these expressions with $M_{b} \approx 5.3$ GeV,
one can estimate the branching ratios for $B$ decays
into $J/\psi + X$ and $\psi^{'} + X$ :
\begin{eqnarray}
B(B \rightarrow J/\psi + X) = 0.23 \%, ~~~~~(0.80 \pm 0.08) \%,
\\
B(B \rightarrow \eta_{c} + X) = 0.14 \%, ~~~~~( < 0.9 \% ~~(90 \% C.L.)),
\\
B(B \rightarrow \psi^{'} + X) = 0.08 \%.~~~~~(0.34 \pm 0.04 \pm 0.03) \%.
\end{eqnarray}
The recent data from CLEO \cite{cleo} are shown in the parentheses, where
the cascades from $B \rightarrow \chi_{cJ}(1P) + X$ followed by $\chi_{cJ}
\rightarrow J/\psi + \gamma$ have been subtracted in the data shown.
In view of these results, we may conclude
there are some important pieces missing in the calculations of decay rates
for $B \rightarrow (c\bar{c})_{1} (^{3}S_{1}) + X$ using the color-singlet
model in the nonrelativistic limit.

In Ref.~\cite{bodwinB}, it was noticed that the inclusion of color-octet
piece, which is often neglected in the previous studies of the
charmonium production in $B$ decays, is mandatory in order to factorize
the amplitude consistently without any infrared divergence in case of
$B$ decays into the $P-$wave charmonium.   This also leads to nonvanishing
decay rates for $B \rightarrow (h_{c}, \chi_{c0}, \chi_{c2}) + X$, all of
which vanish in the color-singlet model.

In view of this, we first estimate the color-octet contributions
to $B \rightarrow
J/\psi +X$, motivated by the suggestion that the color-octet mechanism
might be the solution to the $\psi^{'}$ puzzle at the Tevatron.
Although it is of higher order  in $v^2$ ($\sim O(v^4)$),
it can be important in the case of the inclusive $B$ decays into
$J/\psi + X$, since the Wilson
coefficient of the color-singlet part is suppressed compared to that of the
color-octet part by a factor of $\sim \alpha_{s}$.
( In Eq.~(2), $(2C_{+} - C_{-}) \approx 0.4$, and
$(C_{+} + C_{-} ) \approx 2.20$. )
Now, it is straightforward to calculate the contribution of $(\bar{c} c)_{8}
(^{3}S_{1})$ to $B \rightarrow J/\psi +X$ :
\begin{eqnarray}
\Gamma ( B \rightarrow (\bar{c} c)_{8}(^{3}S_{1}) + X \rightarrow J/\psi+X)
& = & { \langle 0 | O_{8}^{J/\psi} (^{3}S_{1}) | 0 \rangle \over 2 M_{c}^2}~
(C_{+} + C_{-})^{2}~\left( 1 + {8 M_{c}^{2} \over
M_{b}^2} \right)~\hat{\Gamma}_0,
\\
\Gamma ( B \rightarrow (\bar{c} c)_{8}(^{1}S_{0}) + X \rightarrow \eta_{c}
+ X) & = & { 3~\langle 0 | O_{8}^{\eta_{c}} (^{1}S_{0}) | 0 \rangle ~
\over 2 M_{c}^2}~
(C_{+} + C_{-})^{2}~
{}~\hat{\Gamma}_0.
\end{eqnarray}
(Similar expression holds for the $B \rightarrow \psi^{'}
({\rm or}~\eta^{'}) + X$ except that
$\langle 0 | O_{8}^{H} (^{3}S_{1}) | 0 \rangle$ and $M_{c} / M_{b}$ should
change appropriately in order to account for the phase space effects.)

Here, a new nonperturbative parameter $\langle 0 | O_{8} (^{3}S_{1})
| 0 \rangle $ comes in, 
which creates a $(\bar{c} c)$ pair in the color-octet state,
projects into the subspace of states which contain $J/\psi$ in the
asymptotic  future, and then annihilates the  $(\bar{c} c)$ pair at the
creation point.  The matrix element of this operator
is proportional to  the probability that the
color-octet $(\bar{c} c)_{8} (^{3}S_{1})$ to fragment into the physical
$J/\psi$ state in the long distance scale.
This type of a color-octet operator was first considered in
the gluon fragmentation
function  into $P-$wave charmonia in Ref.~\cite{gtochi}, and then  in
the gluon fragmentation into $\psi^{'}$ to solve the $\psi^{'}$ puzzle at
the Tevatron.  Braaten and Fleming fixed $\langle 0 | O_{8}^{\psi^{'}}
(^{3}S_{1}) | 0 \rangle$ to be $4.2 \times 10^{-3}~{\rm GeV}^{3}$
(for $M_{c} \approx 1.5$ GeV), in order to fit the
total cross section for the inclusive $\psi^{'}$ production cross section
at the Tevatron, and found that this value of $\langle 0 | O_{8}^{\psi^{'}}
(^{3}S_{1}) | 0 \rangle$ yields the $p_T$ spectrum
for the $\psi^{'}$ production which nicely agrees with the measured shape.
Then, Cho and Leibovich have performed a complete analysis for the
color-octet contribution to Upsilon and Psi productions at the
Tevatron for both low and high $p_T$ regions \cite{cho}.
Their results are
\begin{eqnarray}
\langle 0 | O_{8}^{J/\psi}  (^{3}S_{1}) | 0 \rangle & = & 1.2 \times
10^{-2}~{\rm GeV}^{3},
\\
\langle 0 | O_{8}^{\psi^{'}}  (^{3}S_{1}) | 0 \rangle & = & 7.3 \times
10^{-3}~{\rm GeV}^{3}.
\end{eqnarray}
Taking the ratio between the color-octet and the color-singlet
contributions,  one gets
\begin{eqnarray}
{\Gamma(B \rightarrow (c\bar{c})_{8}(^{3}S_{1})+X \rightarrow H +X)
\over
\Gamma(B \rightarrow (c\bar{c})_{1}(^{3}S_{1})+X  \rightarrow H +X) }
& = & {3 \langle 0 | O_{8}^{H} (^{3}S_{1}) | 0 \rangle \over 2 \langle 0 |
O_{1}^{H} (^{3}S_{1}) | 0 \rangle}
{}~{(C_{+} + C_{-} )^{2} \over (2 C_{+} - C_{-})^2}
\\
& = & 0.76~~(1.14) {\rm~ for}~ H = J/\psi~~ (\psi^{'}),
\end{eqnarray}
Thus, we find that the color-octet $(c\bar{c})_{8}(^{3}S_{1})$
contributions to
$B \rightarrow J/\psi~~ ({\rm or}~~ \psi^{'}) + X$ are about 76 \%
(114 \%) of the color-singlet contributions in the nonrelativistic limit.

There is another subprocess in the lower order in
$v^2$ compared to the color-octet contribution considered above :
the relativistic corrections to the color-singlet component which is an
order  of $O(v^{2})$ compared to (4),(5).
Extending the Feynman rule in the presence of a heavy quarkonium
\cite{keung},
we can derive that the relativistic corrections to $B \rightarrow J/\psi
{}~({\rm or}~\eta_{c}) +X$
can be written as the following factorized form :
\begin{eqnarray}
\Gamma ( b \rightarrow J/\psi + X) & = &
- { \langle 0 | P_{1}^{J/\psi} (^{3}S_{1}) | 0 \rangle \over 9 M_{c}^4}~
\hat{\Gamma}_{1}
(b \rightarrow (c \bar{c})_{1}~(^{3}S_{1}) + X),
\\
\Gamma ( b \rightarrow \eta_{c} + X) & = &
- { \langle 0 | P_{1}^{\eta_c} (^{1}S_{0}) | 0 \rangle \over M_{c}^4}~
\hat{\Gamma}_{1}
(b \rightarrow (c \bar{c})_{1}~(^{1}S_{0}) + X).
\end{eqnarray}
Here again, one can use the vacuum saturation approximation, Eq.~(1) :
\begin{equation}
\langle 0 | P_{1}^{H} (^{2S+1}S_{J}) | 0 \rangle \approx
(2J+1)~ \langle H | P_{1} (^{2S+1}S_{J}) | H \rangle.
\end{equation}
The latter is related with the spin-weighted average of the $S-$wave
heavy quarkonium wavefunctions  in the following way :
\begin{equation}
\langle nS | P_{1} | nS \rangle = -{3 {\rm Re}~
(\overline{R_{nS}^{*}}~\overline{\nabla^{2} R_{nS})}
\over 2 \pi }~\left( 1 + O(v^{2}) \right),
\end{equation}
$\overline{R_{nS}}$ is the spin-weighted average of the $S-$wave
wavefunctions \cite{bodwin} :
\begin{equation}
\overline{R_{nS}} \equiv {1\over 4}~\left( 3 R_{\psi} + R_{\eta_c} \right).
\end{equation}
In order to estimate the relativistic corrections, we need one more
nonperturbative matrix element, $\langle 0 | P_{1}^{nS} (^{3}S_{1})
| 0 \rangle$ or ${\rm Re}~(\overline{R_{nS}^{*}}
{}~\overline{\nabla^{2}
R_{nS}})$.  This is not available in the current literature now, and we
will determine this parameter as well as $\langle 0 | O_{1} (^{3}S_{1})
| 0 \rangle$ in the following section.
However, we note that the relativistic corrections make the decay rates for
the $S-$wave charmonium productions in $B$ meson decays decrease
because $\langle 0 | P_{1}^{nS} (^{3}S_{1}) | 0 \rangle > 0$.

 \section{Numerical analysis}
 \label{sec:three}

\subsection{ Analysis of the $S-$wave charmonium decays}
\label{subsec:31}

In the previous section, we derived the $S-$wave charmonium production
rates in $B$ decays to $O(v^{2})$ for the color-singlet contributions,
including one of the color-octet contributions at  $O(v^{4})$.
The results depends on three  nonperturbative parameters,
$\langle 0 | O_{1} (^{3}S_{1}) | 0 \rangle, \langle 0 | P_{1}
 (^{3}S_{1}) | 0 \rangle, \langle 0 | O_{8} (^{3}S_{1}) | 0 \rangle$ or
equivalently, $|R_{\psi}(0)|^{2} $ and
${\rm Re} ( \overline{R_{S}^{*}}
\overline{\nabla^{2} R_{S}})$ for the first two color-singlet matrix
elements.
Since the third parameter $\langle 0 | O_{8} (^{3}S_{1}) | 0 \rangle$
is fixed from the fit to the $\psi^{'}$
production at the Tevatron \cite{braaten}, we consider the other two
parameters in this subsection.    In  order to determine these two
parameters, one has to invoke the lattice calculations, some potential
models. Or, one can simply determine these parameters from the well
measured decay rates of $J/\psi,\eta_c$ which depend on the same parameters.
We choose the last method to fix two nonperturbative parameters,
$\langle 0 | O_{1} (^{3}S_{1}) | 0 \rangle,
\langle 0 | P_{1} (^{3}S_{1}) | 0 \rangle$, in this work.  For this
purpose, we list the decay rates for $\psi \rightarrow ggg +
gg\gamma \rightarrow LH ({\rm light ~hadrons})$ and $\psi \rightarrow
l^+ l^-$ \cite{bodwin} :
\begin{eqnarray}
\Gamma ( \psi \rightarrow LH) & = & 2~{\rm Im} f_{1}(^{3}S_{1})
{}~\langle J/\psi | O_{1}(^{3}S_{1}) | J/\psi \rangle
+ 2~ {\rm Im} g_{1}(^{3}S_{1})
{}~\langle J/\psi | P_{1}(^{3}S_{1}) | J/\psi \rangle
+ O(v^{4} \Gamma),
\\
\Gamma ( \psi \rightarrow e^+ e^-) & = & 2~ {\rm Im} f_{ee}(^{3}S_{1})
{}~\langle J/\psi | O_{1}(^{3}S_{1}) | J/\psi \rangle
+ 2~ {\rm Im} g_{ee}(^{3}S_{1})
{}~\langle J/\psi | P_{1}(^{3}S_{1}) | J/\psi \rangle
+ O(v^{4} \Gamma).
\end{eqnarray}
We need to know the short distance coefficients ${\rm Im} f$'s
to $O(\alpha_{s} )$, and ${\rm Im} g$'s to the leading order only in
$\alpha_s$, because of the velocity counting rule in the NRQCD
\cite{bodwin}.   From the results in the earlier
literatures \cite{bodwin} \cite{keung},
one can extract
\begin{eqnarray}
{\rm Im} f_{1} & \equiv & {\rm Im} f_{3g} (J/\psi \rightarrow ggg) +
{\rm  Im} f_{1 \gamma}
(J/\psi \rightarrow  gg\gamma),
\\
{\rm Im} f_{3g}(^{3}S_{1}) & = & {(\pi^{2} - 9)(N_{c}^{2} - 4) C_{F}
\over 54 N_c}~
\alpha_{s}^{3}(M)~
\nonumber
\\  & & \times  \left[ 1 + (-9.46(2)C_{F} + 4.13(17) C_{A} - 1.161(2)
n_{f} ) {\alpha_{s} \over \pi} \right]
\\
{\rm Im} g_{3g}(^{3}S_{1}) & = & - 4 \times 4.33~{(\pi^{2} - 9)(N_{c}^{2}
- 4) C_{F} \over 54 N_c}~\alpha_{s}^{3}(M),
\\
{\rm Im} f_{1 \gamma}(^{3}S_{1}) & = & {2 (\pi^{2} - 9) C_{F}
Q^{2} \alpha \over
3 N_c}~\alpha_{s}^{2}(M) ~\left[ 1 + (-9.46C_{F} + 2.75 C_{A} - 0.774 n_f )~
{\alpha_{s} \over \pi} \right]
\\
{\rm Im} g_{1 \gamma}(^{3}S_{1}) & = & - 4 \times 4.33~{2 (\pi^{2} - 9) C_{F}
Q^{2} \alpha \over 3 N_c}~\alpha_{s}^{2}(M),
\\
{\rm Im} f_{ee}(^{3}S_{1}) & = & {\pi Q^{2} \alpha^{2} \over 3}~
\left[ 1 - 4 C_{F}
{\alpha_{s} \over  \pi} \right],
\\
{\rm Im} g_{ee}(^{3}S_{1}) & = & - {4\over 3}~{\pi Q^{2} \alpha^{2} \over 3},
\\
{\rm Im} f_{qq}(^{3}S_{1}) & = & \pi Q^{2} \left( \Sigma_{i} Q_{i}^{2}
\right) ~\alpha^{2} ~\left[
1 - {13 \over 4 }~C_{F} {\alpha_{s} \over \pi} \right],
\\
{\rm Im} g_{qq}(^{3}S_{1}) & = & - {4\over 3}~\pi Q^{2} \left( \Sigma_{i}
Q_{i}^{2} \right) ~\alpha^{2}.
\end{eqnarray}
In the above expressions, $N_{c} = C_{A} = n_{f} = 3$ and $C_{F} = 4/3$, and
$Q$ is the electric charge of a heavy quark in the unit of the proton charge.
The strong coupling constant $\alpha_{s}(M)$ is defined in the modified
minimal subtraction scheme ($\overline{\rm MS}$) for QCD with $n_f$ light
quarks, renormalized at the scale $\mu = M$.
The last two are relevant to $J/\psi \rightarrow \gamma^{*} \rightarrow$
hadrons.

Similar expressions for the $\eta_{c}$ decays are
\begin{eqnarray}
\Gamma ( \eta_{c} \rightarrow LH) & = & 2~ {\rm Im} f_{1}(^{1}S_{0})
{}~\langle \eta_{c} | O_{1}(^{1}S_{0}) | \eta_{c} \rangle
+ 2~ {\rm Im} g_{1}
(^{1}S_{0}) ~\langle \eta_{c} | P_{1}(^{1}S_{0}) | \eta_{c} \rangle
+ O(v^{4} \Gamma),
\\
\Gamma ( \eta_{c} \rightarrow \gamma \gamma ) & = &
2~ {\rm Im} f_{\gamma \gamma }(^{1}S_{0})
{}~\langle \eta_{c} | O_{1}(^{1}S_{0}) | \eta_{c} \rangle
+ 2~ {\rm Im}
g_{\gamma \gamma}(^{1}S_{0})
{}~\langle \eta_{c} | P_{1}(^{1}S_{0}) | \eta_{c} \rangle
+ O(v^{4} \Gamma),
\end{eqnarray}
where
\begin{eqnarray}
{\rm Im} f_{1}(^{1}S_{0}) & = & {\pi C_{F} \over 2 N_c}~\alpha_{s}^{2}
(M)~\nonumber  \\
& & \left[  1 + \left\{ ~\left( {\pi^2 \over 4} - 5 \right) C_{F} +
\left( {199 \over 18} - {13 \pi^{2} \over 24}~\right) C_{A}
- {8 \over 9}~n_{f} \right\}~{\alpha_{s} \over \pi}~\right],
\\
{\rm Im} g_{1}(^{1}S_{0}) & = & - {4\over 3}~{\pi C_{F} \over 2 N_c}~
\alpha_{s}^{2} (M),
\\
{\rm Im} f_{\gamma \gamma}(^{1}S_{0}) & = & \pi Q^{4} \alpha^{2} ~
\left[ 1 + \left( {\pi^{2} \over 4} - 5 \right) ~ C_{F}
{\alpha_{s} \over  \pi} \right],
\\
{\rm Im} g_{\gamma \gamma}(^{1}S_{0}) & = & - {4\over 3}~\pi Q^{4}
\alpha^{2}.
\end{eqnarray}
The PDG lists the measured data for these decays as follows \cite{pdg} :
\begin{eqnarray}
\Gamma ( \psi \rightarrow LH) & = & ( 60.72 \pm 1.72)~~{\rm keV},
\\
\Gamma ( \psi \rightarrow e^+ e^-) & = & (5.26 \pm 0.37)~~{\rm keV},
\\
\Gamma (\eta_{c} \rightarrow LH) & = & 10.3_{-3.4}^{+3.8}~~{\rm MeV},
\\
\Gamma (\eta_{c} \rightarrow \gamma \gamma) & = & 7.0_{-1.7}^{+2.0}~~
{\rm keV}.
\end{eqnarray}
Since the data on $\eta_{c}$ decays are not precise enough yet, we use
the data on $J/\psi$ decays only in order to determine two parameters
$\langle J/\psi | O_{1}(^{3}S_{1}) | J/\psi \rangle$ and
$\langle J/\psi | P_{1}(^{3}S_{1}) | J/\psi \rangle$.
Since we don't have enough inputs available at this level, we use
$\alpha_{s}(M)$ to be $0.25 \sim 0.28$ instead of treating it as a free
parameter.  Although this choice is not fully systematic from the view
point of perturbative QCD, our numerical results presented below should
not strongly depend on the exact value of $\alpha_{s} (M)$. From
$\psi \rightarrow LH$ and $\psi \rightarrow e^+ e^-$, we determine
\begin{eqnarray}
 \langle J/\psi
 | O_{1} (^{3}S_{1}) | J/\psi \rangle & \approx  &
0.440~(0.490)~{\rm GeV}^{3},
\\
  \langle J/\psi | P_{1} (^{3}S_{1}) | J/\psi \rangle
&  \approx  & 0.025~(0.031)~{\rm GeV}^{5},
\end{eqnarray}
for $\alpha_{s} = 0.25 ~(0.28)$, respectively. For $\psi^{'}$, we get
\begin{eqnarray}
\langle \psi^{'} | O_{1} (^{3}S_{1}) | \psi^{'} \rangle
& \approx & 0.177~(0.198)~~{\rm GeV}^{3},
\\
 \langle \psi^{'} | P_{1} (^{3}S_{1}) | \psi^{'} \rangle
& \approx & 0.008 ~(0.011)~~{\rm GeV}^{5}.
\end{eqnarray}
Note that the radiative corrections to the $J/\psi$ and $\psi^{'}$
decays are fairly large with or without the $\langle 0 | P_{1}
(^{3}S_{1}) | 0 \rangle$ term.
Radiative corrections increase $\langle J/\psi | O_{1} (^{3}S_{1}) | J/\psi
\rangle$ in Eq.~(9) by $\sim 80 \%$.
The relativistic correction term, $\langle J/\psi | P_{1} (^{3}S_{1})
| J/\psi \rangle$, is
about $\sim 9 \%$ of Eq.~(9), hence decreases the $B
\rightarrow J/\psi + X$ rate (Eq.~(4)) by $\sim 2 \%$.

Next, let us determine $\langle \eta_{c} | O_{1}(^{1}S_{0}) | \eta_{c}
\rangle$  which is expected to be
\begin{equation}
\langle \eta_{c} | O_{1}(^{1}S_{0}) | \eta_{c} \rangle =
\langle J/\psi | O_{1}(^{3}S_{1}) | J/\psi \rangle
{}~(1 + O(v^2)),
\end{equation}
due to the heavy quark spin symmetry.
Since $\langle 0 | P_{1} (^{3}S_{1}) | 0 \rangle$ is independent
of the total spin $S$, one can determine
$\langle \eta_{c} | O_{1}(^{1}S_{0}) | \eta_{c} \rangle$
from one of the decays, Eq.~(38) or Eq.~
(39), and then predict the other and compare with the measured rate.
Also, the relation (52) should be respected in order to be consistent with
the NRQCD and the heavy quark  spin symmetry.
If we use (46) as an input, we get (with $\alpha_{s} = 0.25$)
\begin{eqnarray}
\langle \eta_{c} | O_{1}
 (^{1}S_{0}) | \eta_{c} \rangle
& \approx  & (0.149^{+0.053}_{-0.046})~{\rm GeV}^{3},
\\
\Gamma ( \eta_{c} \rightarrow \gamma \gamma ) & = & (2.8^{+1.1}_{-1.0})
 ~{\rm keV},~~
\end{eqnarray}
the former of which severely violates the heavy quark spin symmetry,
(52). For $\alpha_{s} = 0.28$, the numbers above change into
$\left( 0.116^{+0.039}_{-0.035} \right)~{\rm GeV}^{3}$
and $\left( 1.8^{+ 0.8}_{- 0.7  } \right)$ keV.
On the other hand, if we use (47)
as an input with with $\alpha_{s} = 0.25$, we get
\begin{eqnarray}
\langle \eta_{c} |
O_{1} (^{1}S_{0}) | \eta_{c} \rangle & \approx  &
(0.326^{+0.092}_{-0.080})~{\rm GeV}^{3},
\\
\Gamma ( \eta_{c} \rightarrow LH ) & = & ( 23^{+7}_{-6})
 ~{\rm MeV},
\end{eqnarray}  For $\alpha_{s} = 0.28$, the numbers become
$\left( 0.341^{+0.097}_{-0.083} \right)~{\rm GeV}^3$,
and $\left( 32^{+9}_{-8 } \right)$  MeV,
respectively.
Now, the relation (48) is better obeyed,  although the difference between
$\langle J/\psi | O_{1} (^{3}S_{1}) | J/\psi  \rangle$ and $\langle
\eta_{c} | O_{1} (^{1}S_{0}) |  \eta_{c} \rangle$ is not that small,
 and the predicted
rate for $\eta_{c} \rightarrow $LH is quite large compared to the data,
(46).
One may conclude that the factorization formulation by BBL in terms of the
NRQCD predicts rather large value of $\eta_{c} \rightarrow $LH, compared
to the current experimental values, considering the large uncertainties in
the measurements. A better determination of $\Gamma ( \eta_{c} \rightarrow
{}~{\rm LH})$ would test the validity of the factorization approach to
$O(v^2)$.

\subsection{Results for $B$ decays}
\label{subsec:32}

Since all the relevant nonperturbative parameters are in hand  now, we are
ready to estimate  the branching ratio for $B \rightarrow J/\psi + X$
which includes $O(v^2)$ corrections and one of the $O(v^4)$ color-octet
contribution.  Adding up the change in $\langle 0 | O_{1}^{H}(^{3}S_{1})
| 0 \rangle $ and the color-octet
contributions, we get a moderate increase in the branching ratio by factor
of $\approx (1.0 + 0.80 - 0.02 + 0.76) = 2.54$ of the lowest order
prediction (13) for the $B \rightarrow J/\psi
+ X$ case, and $(1.0 + 0.80 - 0.02 + 1.14) = 2.92$ for the $\psi^{'}$
case :
\begin{eqnarray}
B(B \rightarrow J/\psi + X) & = & 0.58~ \%,
\\
B(B \rightarrow \psi^{'} + X) & = & 0.23~ \%,
\end{eqnarray}
compared to the data, $(0.80 \pm 0.08) \%$ and $(0.34 \pm 0.04 \pm 0.03)
\% $.  We note that the agreements between theoretical estimates and the
data get improved, after the radiative corrections and the color-octet
mechanism  have been included.  There is a residual uncertainty related
with the $b$ quark mass $M_{b}$. In this work, we have chosen $M_{b}
\approx 5.3$ GeV,  and normalized the decay rate to that of the
semileptonic  $B$ decay in order to reduce the uncertainty from less
known $M_b$ \cite{bodwinB}.  If we use $M_{b} = 4.5 $
GeV, for example, all the decay rates should be multiplied by a factor of
$5.3/4.5 \approx 1.18$.   For $B \rightarrow \eta_{c} + X$,
our prediction is tampered by less known
$\langle 0 | O_{1}^{\eta_{c}} (^{1}S_{0}) | 0 \rangle$ as well as
$\langle 0 | O_{8}^{\eta_{c}} (^{1}S_{0}) | 0 \rangle$.
Still, we expect that the decay rate increases by
a factor of $\sim 2$ or more over the lowest order result, (14).

Let us finally consider the polarization of $J/\psi$'s produced in
$B$ decays.  It is convenient to define two parameters, $\zeta$ and
$\alpha$ as follows :
\begin{eqnarray}
\zeta  & =  & {\Gamma_{T} (B \rightarrow J/\psi + X) \over
\Gamma_{T+L}  (B \rightarrow J/\psi + X) },
\\
\alpha  & = & {{3 \zeta - 2} \over {2 - \zeta}}.
\end{eqnarray}
The quantity $\alpha$ can be readily measured through the polar angle
distribution of dileptons in $J/\psi \rightarrow l^{+} l^{-}$ in the rest
frame of $J/\psi$ :
\begin{equation}
{{d\Gamma (J/\psi \rightarrow l^{+} l^{-})} \over d \cos \theta}
\propto 1 + \alpha \cos^{2} \theta,
\end{equation}
where $\theta$ is the angle between the flight direction of a lepton
in the rest frame of $J/\psi$ and the flight direction
of $J/\psi$ in the rest frame of the initial $B$ meson.  For the
unpolarized $J/\psi$, we would have $\zeta = 2/3 ~(\alpha = 0)$
corresponding to the flat $\cos \theta$ distribution of dileptons.
Assuming the factorization for $B \rightarrow J/\psi + X$ in the
color-singlet  component in the effective Hamiltonian (2), one finds that
$J/\psi$'s  produced in $B$ decays are  polarized with \cite{falk}
\begin{equation}
\alpha = - {{M_{b}^{2} - M_{\psi}^2} \over  {M_{b}^{2} + M_{\psi}^2}}
\approx  - 0.49~~~(-0.36)
\end{equation}
or $\zeta = 0.41 ~~(0.49) $ for $M_{b} = 5.3~~(4.5)$ GeV,
which nicely compares with the CLEO measurement \cite{cleoold},
$\alpha_{exp} =-0.44 \pm 0.16$.
One may wonder if the color-octet mechanism considered in this work
can change
the polarization of $J/\psi$ substantially.  However, the structure of the
amplitude  for $ b \rightarrow J/\psi + X$ due to the color-octet $^{3}S_1$
state is the same as that due to the color-singlet mechanism (including both
the lowest and the next-to-leading order terms in $v^2$).  Namely, the
amplitude for $b \rightarrow J/\psi + X$ is proportional to
\begin{equation}
{\cal M} \propto \epsilon_{\mu}~\bar{s} \gamma^{\mu} (1-\gamma_{5}) b.
\end{equation}
Thus,
the prediction for $\alpha$ remains the same even in the presence of  the
$(\bar{c}c)_{8}(^{3}S_{1})$ color-octet contribution.
This is in sharp contrast with the case of
$\psi^{'}$ production through the gluon fragmentation into the color-octet
$c\bar{c}$ state which in turn evolves into $\psi^{'}$. In the case of
gluon fragmentation ($g \rightarrow (c\bar{c})(^{3}S_{1})_{8} \rightarrow
J/\psi + X$), the initial gluon (with $q^{2} \approx (2 M_{c})^2$ and
$q^{0} >> 2 M_{c}$)  is almost on shell, being almost
transverse up to $q^{2} / q_{0}^2$.  Thus, the polarization of the
color-octet  $c\bar{c}$ is also almost transverse.  Because of the spin
symmetry of heavy quark system, the polarization of the daughter $J/\psi$
is the same as the  parental color-octet $c\bar{c}$ state.

 \section{Conclusions}
 \label{sec:con}

In this work, we considered the relativistic corrections and a color-octet
contribution to $S-$wave charmonium  productions in $B$ decays.
Our results, (57) and (58), give moderate increase to the previous analyses
based on the color-singlet mechanism in the nonrelativistic limit.
Compared to the previous analyses of the lowest order in $v^2$ and
$\alpha_s$,  we get an $\sim 80 \%$ increase in $\langle J/\psi | O_{1}
(^{3}S_{1}) | J/\psi \rangle$ from the radiative corrections to
$J/\psi \rightarrow e^{+} e^{-}$ and $J/\psi \rightarrow $ LH, $\sim  2\%$
decrease from the relativistic correction through the $\langle J/\psi
| P_{1} (^{3}S_{1}) | J/\psi \rangle$ term, and
$\sim 76 \%$ increase in the decay rate from the color-octet contribution,
$\langle 0 | O_{8}^{J/\psi}(^{3}S_{1}) | 0 \rangle$.  Thus, the color-octet
mechanism, which has been proposed as a possible solution to the $\psi^{'}$
puzzle at  the Tevatron, could give an enhancement of the
$B \rightarrow J/\psi + X$ decay  rate by a moderate amount.

It should be kept in mind that  what we considered
in this work is only one of the  color-octet operators that may contribute
to $B \rightarrow J/\psi + X$.
We have chosen only $\langle 0 | O_{8}^{J/\psi}
(^{3}S_{1}) | 0 \rangle$, since we know the numerical value of this matrix
element from the work of Braaten and Fleming \cite{braaten}.  This matrix
element is rather special in the sense that it is the only color-octet
operator which is relevant to $g \rightarrow (c\bar{c})_{8} (^{3}S_{1})
\rightarrow J/\psi + X$ in the leading order in $v^2$ and $\alpha_s$.
However, for the $B$ decays, other color-octet operators can contribute as
well ; e.g.,
\begin{eqnarray}
\Gamma ( B \rightarrow (c\bar{c})_{8}(^{1}S_{0}) + X \rightarrow
J/\psi + X ) & = & {3 \langle 0 | O_{8}^{J/\psi}(^{1}S_{0}) | 0 \rangle
\over 2 M_{c}^2}~(C_{+} + C_{-})^{2}~\hat{\Gamma}_{0},
\\
\Gamma ( B \rightarrow (c\bar{c})_{8}(^{3}S_{1}) + X \rightarrow
\eta_{c} + X ) & = & {\langle 0 | O_{8}^{\eta_c}(^{3}S_{1})  | 0 \rangle
\over 2 M_{c}^2}~(C_{+} + C_{-})^{2}~\left( 1 + {8 M_{c}^{2} \over M_{b}^2}
\right)~\hat{\Gamma}_{0},
\end{eqnarray}
and similar expressions for the contributions of $\langle 0 | O_{8}^{H}
(^{3}P_{J}) | 0 \rangle$ :
\begin{eqnarray}
\Gamma ( B \rightarrow (c\bar{c})_{8}(^{3}P_{1}) + X \rightarrow
J/\psi + X ) & = & { \langle 0 | O_{8}^{J/\psi}(^{3}P_{1}) | 0 \rangle
\over  M_{c}^4}~(C_{+} + C_{-})^{2}~\left( 1 + {8 M_{c}^{2} \over M_{b}^2}
\right)~\hat{\Gamma}_{0},
\\
\Gamma ( B \rightarrow (c\bar{c})_{8}(^{3}P_{1}) + X \rightarrow
\eta_{c} + X ) & = & {\langle 0 | O_{8}^{\eta_c}(^{3}P_{1})  | 0 \rangle
\over   M_{c}^4}~(C_{+} + C_{-})^{2}~\left( 1 + {8 M_{c}^{2} \over M_{b}^2}
\right)~\hat{\Gamma}_{0},
\end{eqnarray}
In order to estimate the effects of these color-octet matrix elements
$\langle 0 | O_{8}^{J/\psi}(^{1}S_{0}) | 0 \rangle$ and
$\langle 0 | O_{8}^{J/\psi}(^{3}P_{1}) | 0 \rangle$,
we need to consider other processes as well,
such as $\gamma + p \rightarrow J/\psi +
X$ and $e^{+} e^{-} \rightarrow \gamma^{*} \rightarrow J/\psi + X$
\cite{chen}.   For example,
the height of the elastic peak for photoproduction of $J/\psi$
depends on these color-octet matrix elements \cite{braaten}.
Complete analysis of color-octet
contributions to the $S-$wave charmonia in $\gamma p$ collisions is
called for as well \cite{eptocc}.
Once these new color-octet matrix elements are determined from other
processes, our results in this work will provide an independent test of
the hypothesis of color-octet mechanism as a possible solution to
$\psi^{'}$ anomaly at the Tevatron.

We have also analyzed the leptonic and the inclusive hadronic decays of
$J/\psi$ and $\psi^{'}$ to $O(v^{2})$ in the framework of the
BBL's factorization scheme, and did extract two nonperturbative
parameters, $\langle H | O_{1}(^{3}S_{1}) | H \rangle$ and $\langle H |
P_{1}(^{3}S_{1}) | H \rangle$ with $H = J/\psi$ and $ \psi^{'}$.
These are important inputs in many other
theoretical calculations of the $S-$wave charmonia productions in various
high energy processes and their subsequent decays.
As a by-product, we have found that the inclusive hadronic decay rate for
$\eta_c$ may be larger than the current PDG value by factor of $\sim 2$,
if the BBL's factorization formulae to $O(v^{2})$ works with the charmonium
system.
The better measurements of $\Gamma (\eta_{c} \rightarrow$ LH) would test
our predictions based on the factorization approach for the heavy quarkonium
decays in the framework of NRQCD to $O(v^{2})$.

\acknowledgements
We are grateful to Dr. Seyong Kim for discussions on this subject as well
as the current status of the lattice calculations of the nonperturbative
parameters, $|R_{S}(0)|^2$ and ${\rm Re}(\overline{R_{S}^{*}}
\overline{\nabla^{2} R_{S}})$ in the $\Upsilon$ system.

This work is supported in part by KOSEF through CTP at Seoul National
University. Also,
P.K. was supported in part by NON DIRECTED RESEARCH FUND,
Korea Research  Foundations,
and in part by the Basic Science Research Institute
Program, Ministry of Education, 1994, Project No. BSRI--94--2425.

\end{document}